\documentclass[aps,twocolumn,prb]{revtex4-1}

\usepackage{amsmath,amssymb,bm}
\usepackage{graphicx}
\usepackage{epstopdf}
\usepackage{latexsym}
\usepackage{subfigure}
\usepackage{color}
\usepackage{hyperref}
\usepackage{natbib}
\usepackage{braket}

\newcommand{\rvec}{\mathbf{r}}

\newcommand{\suN}{SU($N$) }
\newcommand{\aniso}{\gamma}
\newcommand{\Neel}{N\'{e}el}

\begin{document}

\title{The N\'{e}el-VBS transition in three-dimensional SU($N$) antiferromagnets}
\author{Matthew S. Block}
\author{Ribhu K. Kaul}
\affiliation{Department of Physics and Astronomy, University of Kentucky, Lexington, Kentucky 40506, USA}

\date{\today}

\begin{abstract}
We present results for the phase diagram of an SU($N$) generalization of the Heisenberg antiferromagnet on a bipartite three-dimensional anisotropic cubic (tetragonal) lattice as a function of $N$ and the lattice anisotropy $\gamma$. In the ``isotropic'' $\gamma=1$ cubic limit, we find a transition from N\'{e}el to valence bond solid (VBS) between $N=9$ and $N=10$. We follow the N\'{e}el-VBS transition to the limiting cases of $\gamma \ll 1 $ (weakly coupled layers) and $\gamma \gg 1$ (weakly coupled chains).    Throughout the phase diagram we find a direct {\em first-order} transition from N\'{e}el at small-$N$ to VBS at large-$N$. In the three-dimensional models studied here, we find no evidence for either an intervening spin-liquid ``photon'' phase or a continuous transition, even close to the limit $\gamma \ll 1$ where the isolated layers undergo continuous N\'{e}el-VBS transitions.

\end{abstract}


\maketitle

\section{Introduction}
\label{sec:intro}

The theoretical and experimental study of the destruction of magnetic order at $T=0$ has become one of the most popular topics of quantum condensed matter physics in recent decades.~\cite{Sachdev11} Such phase transitions arise in a general class of many body systems referred to as ``quantum magnets,'' prominent examples being  transition metal oxide systems, heavy fermion materials, and ultra cold atomic gases in optical lattices. Given the great interest in this phenomena in experimental systems, an extensive body of theoretical work has been devoted to the study of the simplest microscopic models that display these phase transitions, i.e., quantum spin systems. Model systems have been studied theoretically by a large variety of analytic~\cite{Auerbach94,Sachdev11} and numerical methods.~\cite{Stoudenmire12,Sandvik10b,kaul2013:arcmp_arxiv} 

The most popular spin models are anti-ferromagnets with SU(2) symmetry. Later generalizations to SU($N$) symmetry have also received a lot of attention, both in the large-$N$ limit and for finite values of $N$. For bipartite lattice models, the simplest magnetic order is of the semi-classical kind, which we will term N\'eel order for all $N$. On increasing quantum fluctuations, theoretical reasoning suggests that a natural non-magnetic state to consider in the case of many anti-ferromagnets is a translational symmetry breaking - valence bond solid (VBS). A natural line of investigation that then arises is the theoretical description of the N\'eel-VBS transition. Interestingly the nature of this transition is strongly affected by the dimensionality of the system. In this work we study the transition between N\'eel and valence bond solid (VBS) phases in antiferromagnets with SU($N$) symmetry {\em in three spatial dimensions} by unbiased quantum Monte Carlo (QMC) simulations. Before we turn to an exposition of our results, we will briefly review past results of studies of the SU($N$) N\'eel-VBS transition in  one and two dimensions.

\begin{figure}[t]
\centerline{\includegraphics[width=\columnwidth]{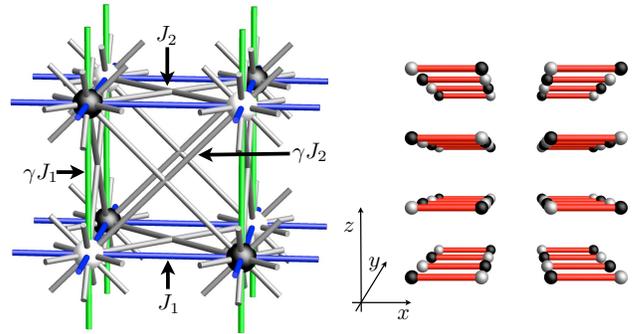}}
\caption{(color online).  Three dimensional lattice on which the model, Eq.~\ref{eqn:model}, is defined. Left: A 2$\times$2$\times$2 piece of the cubic lattice showing nearest neighbor ($J_1$ and $\aniso J_1$) and next-nearest neighbor ($J_2$ and $\aniso J_2$) couplings. Bonds in the $x-y$ plane have strength $J_1,J_2$. The parameter $\aniso$ is a multiplicative factor for all bonds with a $z$-component that connect neighboring $x-y$ planes.  Right: The pattern of red bonds depicts one of the degenerate columnar dimer patterns found in the VBS phase of the model given by Eq.~\ref{eqn:model}. The dimer pattern is such that the Bragg peaks are located at $(\pi,0,0)$ and $(0,\pi,0)$ for $\gamma <1$  and at $(0,0,\pi)$ for $\gamma>1$. At $\gamma=1$ all three Bragg reflections appear.}
\label{fig:latt}
\end{figure}

In {\em one dimension} the transition between ``N\'eel'' and VBS can be studied accurately by controlled theoretical methods.~\cite{giamarchi2004:book} We used the quotation marks because in one dimension there can be no true long-range order for the N\'eel phase, although true long-range order is possible at $T=0$ for the VBS. The transition between the power-law ``N\'eel'' and VBS states has been studied in detail in the anti-ferromagnetic SU(2) $J_1$-$J_2$ model.~\cite{chitra1995:dmrgj1j2} More recently, the same transition has been studied in sign problem free one-dimensional JQ models.~\cite{tang2011:1djq,sanyal2011:1djq} The theoretical expectation of a Kosterlitz-Thouless type transition between ``N\'eel'' and VBS states has been verified in numerical studies.

In {\em two dimensions} the situation gets more interesting since true long range order is possible for the N\'eel state at $T=0$. While conventional Landau theories would predict a first-order transition or an intervening phase, a field theoretic work made the prediction of a direct continuous phase transition between conventional {\Neel} ordered and VBS phases, called a deconfined critical point.~\cite{Senthil04a} There is now strong evidence that the deconfined scenario is realized in a series of models that allow the transition to be studied by QMC simulations as a function of $N$ in SU($N$) antiferromagnets.~\cite{Sandvik07,lou2009:sun,kaul2011:j1j2}
The SU($N$) studies give quantitative estimates for scaling dimensions of both N\'eel and VBS order parameters for large values of $N$. By comparing these universal numbers to the estimates from a $1/N$ expansion~\cite{halperin1974:largeN} of the relevant three-dimensional classical non-compact CP$^{N-1}$ field theory,~\cite{motrunich2004:o3,kamal1993:o3} which had previously been computed analytically,~\cite{kaul2008:u1,murthy1990:mono,metlitski2008:mono} substantial credence was given to the identification of deconfined criticality in this system.

In {\em three dimensions} the nature of the N\'eel-VBS phase transition has been studied only in SU(2) models~\cite{beach2007:3d} with full cubic symmetry, where clear evidence for a first-order transition was reported. In this work we extend the previous study~\cite{beach2007:3d} in two ways of interest: we study the N\'eel-VBS transition in systems that have SU($N$) symmetry with $2\leq N\leq 10$, and we study lattices with anisotropic couplings so that we can study the phase transition in the limit of weakly coupled two-dimensional (2D) layers. The motivation for these generalizations comes from our theoretical understanding of the N\'eel-VBS transition in two dimensions, where deconfined criticality arises due to the instability of three-dimensional (3D) compact gauge theories to confinement. In $(3+1)$ dimensions, no instability exists for a compact gauge theory and thus one might expect to find an exotic gapless ``photon'' phase close to a transition between N\'eel and VBS.

In this work, we endeavor to answer the following questions: If one were to couple together 2D planes, each residing close to the deconfined quantum critical point, would the emergent gapless photon remain deconfined over some finite region of the phase space realizing the eagerly sought after photon phase in three dimensions?  And, if there were no intervening phases between the {\Neel} ordered and VBS phases, what is the nature of the transition between the two phases in three dimensions? The search for a photon phase in microscopic models has been carried out in various contexts. Models in which there is evidence for the photon phase include large-$N$ versions of Sp($N$) magnets,~\cite{bernier2005:u1vbs} dimer models,~\cite{huse2003:u1,sikora2009:qdm_dia} and U(1) symmetric magnets.~\cite{hermele2004:pyrochlore,motrunich2005:u1,banerjee2008:pyro}

In our study here, we will work on a bipartite lattice in which the spins transform as a fundamental representation on one sub-lattice and a conjugate to fundamental on the other sublattice.~\cite{affleck1985:lgN,Read90} The specific lattice considered here may be thought of as an anisotropic cubic lattice (really a tetragonal lattice), where the coupling on the bonds in one direction (say the $z$ direction) is different from the coupling on the bonds in the other two directions ($x$ and $y$ directions). For SU(2) it is well known that independent of the anisotropy of the lattice, this model is N\'eel ordered. On the other hand, when $N$ is very large the SU($N$) antiferromagnet maps to a quantum dimer model with only a kinetic term.~\cite{read1989:nucphysB} This model is valence bond solid ordered at $T=0$ independent of the extent of anisotropy.~\cite{moessner2003:3dqdm} From these simple considerations, it is clear that regardless of the anisotropy there must be a transition between N\'eel and VBS at some finite value of $N$. The central questions we have investigated here are whether there is a new intervening phase between the N\'eel and VBS phases and the nature of the transitions, i.e., whether they are first order or continuous as a function of $N$.

In Sec.~\ref{sec:model}, we define a 3D model, by generalizing the 2D $J_1$-$J_2$ model,~\cite{kaul2011:j1j2} and in Sec.~\ref{sec:mm} we explain the methods by which we simulate our model. In Sec.~\ref{sec:pd} we present the phase diagram inferred from our QMC simulations, and in Sec.~\ref{sec:trans} we study the nature of the phase transitions. Finally, in Sec.~\ref{sec:conc} we summarize our results and present an outlook on future work.

\section{The Model}
\label{sec:model}

\begin{figure}[t]
\centerline{\includegraphics[width=\columnwidth]{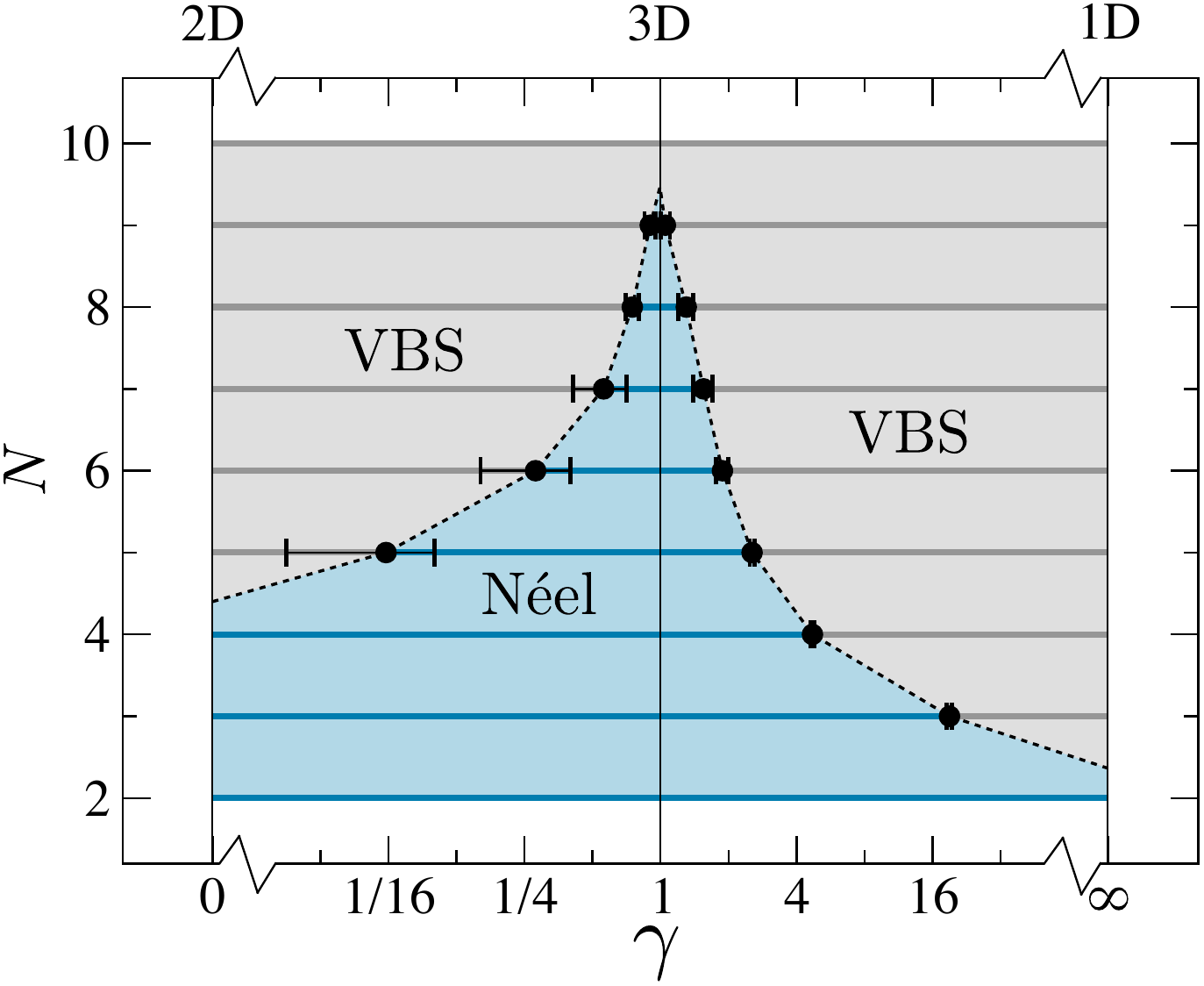}}
\caption{(color online).  Phase diagram of the $J_1$ model, i.e., Eq.~(\ref{eqn:model}) with $J_2=0$, showing the transitions between {\Neel} and VBS order as $\aniso$ is varied for values of $N$ ranging from 2 to 10.  For $N=5,6,7,8,9$, there are three segments for each $N$: the first (gray) corresponds to VBS order, the second (blue) corresponds to {\Neel} order, and the third (gray) is again VBS order.  For $N=2$, the system is always magnetically ordered even in pure 1D (the so-called Bethe-chain phase).  Likewise, for $N=10$, the system is always VBS ordered in three or less dimensions.  Also note that for $N=3,4$, there are only two segments, {\Neel} then VBS, as the system remains {\Neel} ordered on the 2D side of the phase diagram since $N=5$ is required in pure 2D to destroy the magnetic order.  The horizontal axis is shown on a log$_2$ scale. The value of $N$ for which the system VBS orders in the 2D limit~\cite{harada2003:sun,beach2009:sun} and 1D limit~\cite{barber1989:d1n3_vbs} have been studied previously.}
\label{fig:pd}
\end{figure}
We construct our model from simple \suN invariant operations.  We first imagine our system of spins on some bipartite lattice is restricted to a Hilbert space wherein the states on the sites of the $A$ sublattice transform under the fundamental representation of \suN while the states on the sites of the $B$ sublattice transform under the conjugate to the fundamental representation, a convention common in earlier works.~\cite{Read89,harada2003:sun,beach2009:sun} This results in a state of the form $\sum_\alpha\ket{\alpha}_A\ket{\alpha}_B$ being an \suN singlet.  The first simple operation we include in our Hamiltonian is the projector $P_{ij}$ onto this singlet between two sites $i$ and $j$ on different sublattices.  If one takes $H=-J_1/N\sum_{\braket{ij}}P_{ij}$, this reduces to the usual near-neighbor Heisenberg antiferromagnet (up to a constant) for $N=2$.  On the square lattice in two dimensions, it is already known that the {\Neel} order gives way to VBS order for integer $N\geq5$ (Refs.~\onlinecite{harada2003:sun}~and~\onlinecite{beach2009:sun}); hence, the transition occurs between $N=4$ and $N=5$ if $N$ is viewed as a continuous parameter. We expect that on the cubic lattice in three spatial dimensions, there will similarly exist some value of $N$ (presumably larger than 5) for which the {\Neel} order fails to persist. We include a second term of another simple operation that permutes the spins on two sites of the \emph{same} sublattice, $\Pi_{ij}$.  This term is the \suN generalization of the Heisenberg ferromagnet.  With only this second term present, the system will trivially order ferromagnetically on each sublattice for any value of $N$.  Turning on the first, antiferromagnetic term with small coupling will lock in the {\Neel} order.  Therefore, if we start with only the first term and some $N$ sufficiently large such that the magnetic order is destroyed, we can presumably turn on and increase the coupling of the second term until {\Neel} order is restored.  In this way, we can continuously tune to the quantum transition point for each $N$ on the VBS side of the transition.  See Fig.~\ref{fig:latt} for a depiction of the lattice and the relevant couplings.

Finally, we would like to be able to vary the strength of the coupling between the 2D layers so we can interpolate from the results already known in one and two dimensions to the isotropic 3D limit.  We accomplish this with the anisotropy parameter $\aniso$.  Denoting the vector pointing from site $i$ to $j$ as $\vec{r}_{ij}$, we have:
\begin{align}
\label{eqn:model}
H=&-\frac{J_1}{N}\sum_{\substack{\braket{ij}\\\vec{r}_{ij}\cdot\hat{z}=0}}P_{ij}-\aniso\frac{J_1}{N}\sum_{\substack{\braket{ij}\\\vec{r}_{ij}\cdot\hat{z}\neq 0}}P_{ij}\\\nonumber&-\frac{J_2}{N}\sum_{\substack{\braket{\braket{ij}}\\\vec{r}_{ij}\cdot\hat{z}=0}}\Pi_{ij}-\aniso\frac{J_2}{N}\sum_{\substack{\braket{\braket{ij}}\\\vec{r}_{ij}\cdot\hat{z}\neq 0}}\Pi_{ij},
\end{align}
where $\langle ij\rangle$ denotes nearest neighbors and $\langle \langle ij\rangle\rangle$ denotes next-nearest neighbors on the underlying cubic lattice. The limit $\aniso=0$ recovers the 2D case, $\aniso\rightarrow\infty$ gives a system of decoupled one-dimensional (1D) chains, and $\aniso=1$ is the 3D isotropic case.  It is worth mentioning here that the situation in 1D is also well studied; assuming $J_2=0$, only for $N=2$ do we have the analog of {\Neel} order, the so-called Bethe phase; for $N\geq3$, there is again the VBS order~\cite{affleck1985:lgN,barber1989:d1n3_vbs} indicating a transition between $N=2$ and $N=3$.  One of the goals of this paper is to determine which integer values of $N$ correspond to {\Neel} and VBS order as we vary $\aniso$ from zero to infinity. The coupling $J_2$ is introduced so that we can tune continuously across the N\'eel-VBS phase transition at fixed values of $N$ and $\gamma$.

\begin{figure}[t]
\centerline{\includegraphics[width=\columnwidth]{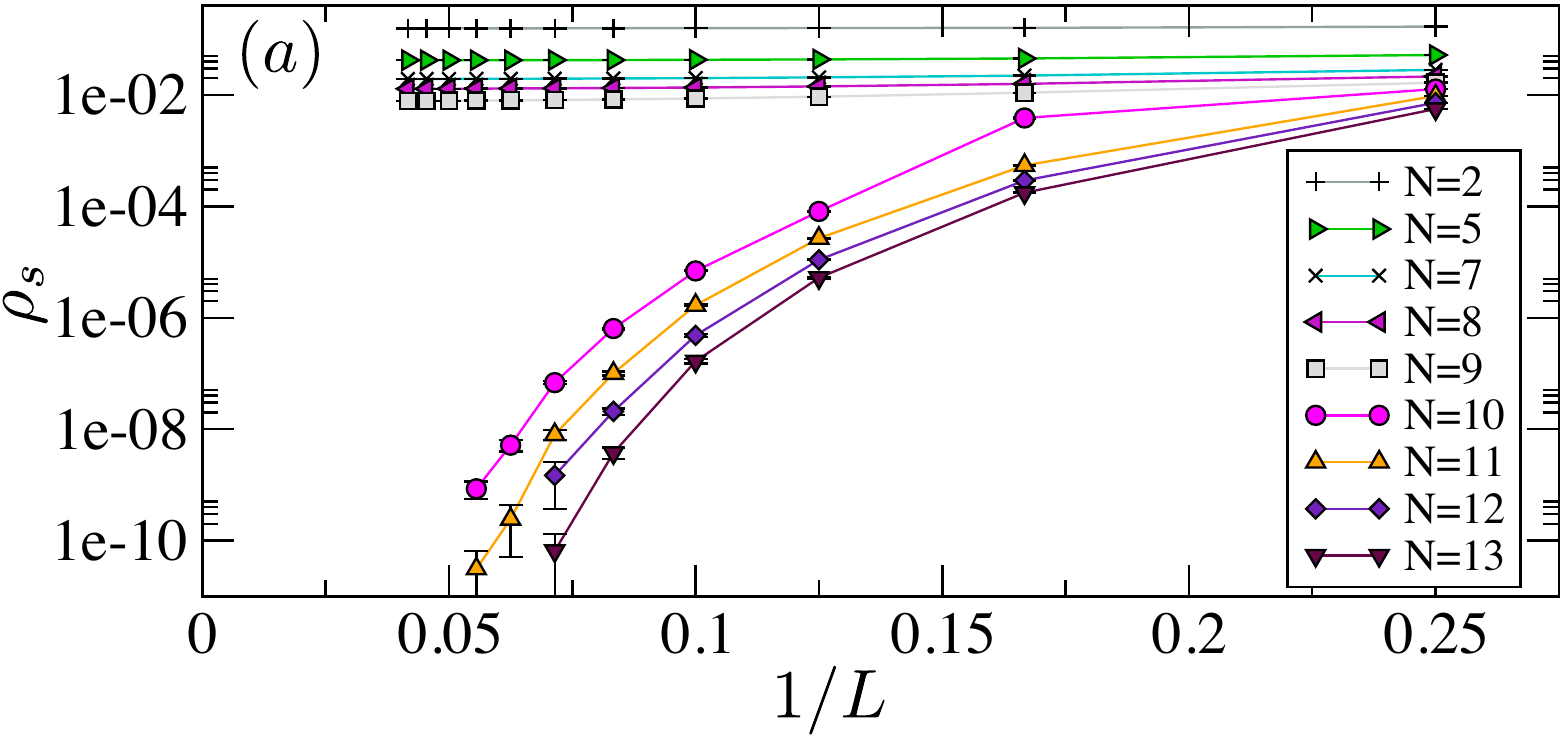}}
\centerline{\includegraphics[width=\columnwidth]{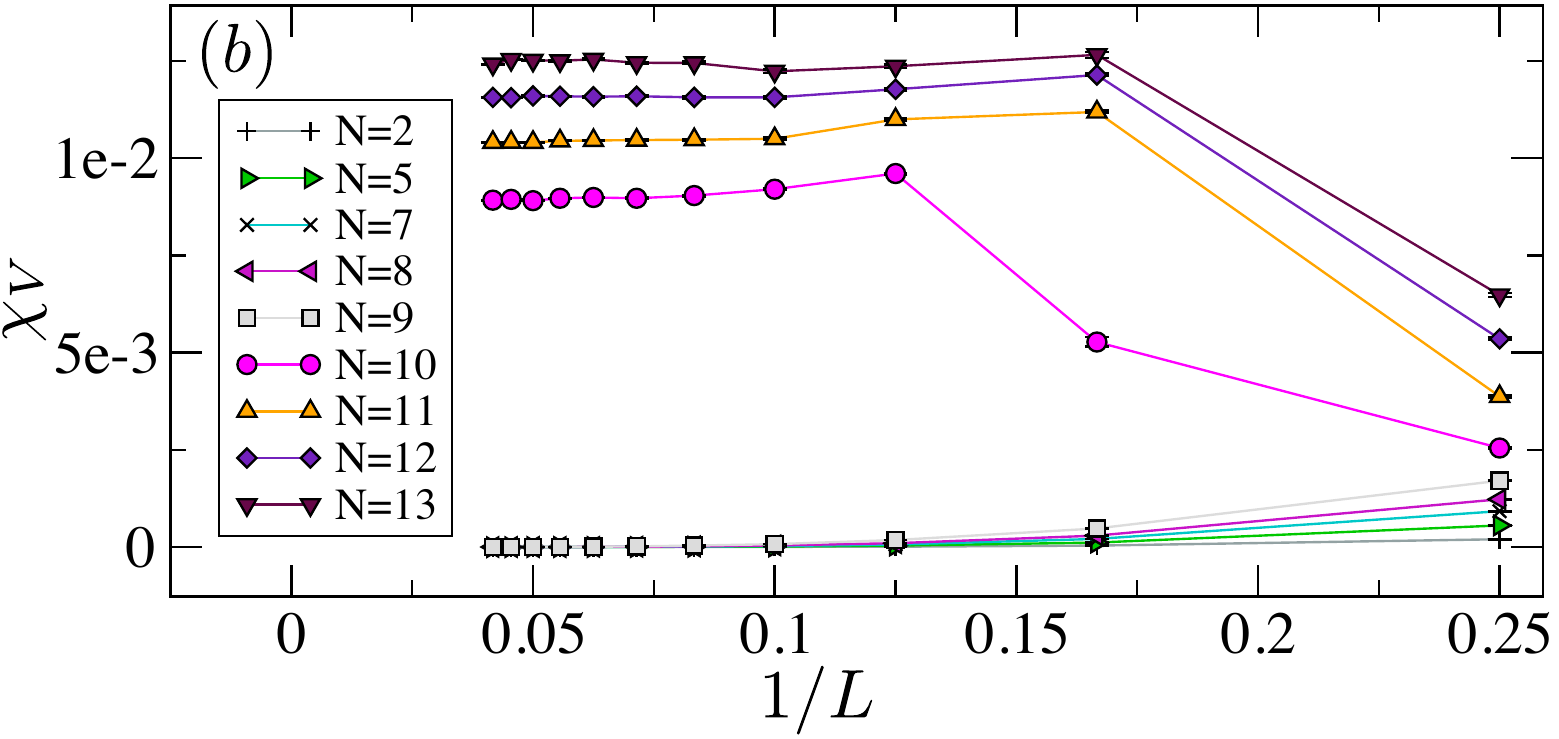}}
\caption{(color online).  Plots demonstrating the disappearance of magnetic order and onset of VBS order. The data shown here are for $\aniso=1$.  (a) Spin stiffness, $\rho_s$ (log$_{10}$ scale), vs. $1/L$ for various values of $N$. (b) Square of VBS order parameter, $\chi_V$, vs. $1/L$ for various values of $N$.}
\label{fig:isoN}
\end{figure}

\section{Methods and Measurements}
\label{sec:mm}

All of our measurements were obtained exactly (to within statistical error) using the stochastic series expansion QMC method with global loop updates.~\cite{Sandvik99,Evertz03,Sandvik10b} The model presented in the previous section explicitly has all non-positive matrix elements leading to positive weights in the partition function and therefore no sign problem.  Finite size scaling is performed on each quantity using a series of system sizes that varies depending on the anisotropy parameter.  The system size is always taken to be $L\times L\times L_z$ where the ratio $L/L_z$ is adjusted to accommodate the anisotropy in a manner that will be explained below.

{\Neel} order is detected via the measurement of a finite spin stiffness $\rho_s$, which is determined from the fluctuations of the spatial winding number of the world lines, $W_\mu$. Specifically, $\beta L_\mu\rho_{s,\mu}=\braket{(W_\mu)^2}$, where $\mu=x,y,z$ and $L_x=L_y=L$.  In the isotropic case, we report the average of all three components of the spin stiffness, $\rho_s=(\rho_{s,x}+\rho_{s,y}+\rho_{s,z})/3$ while on the 2D side of isotropy, we only average the $x$ and $y$ spin stiffnesses: $\rho_s=(\rho_{s,x}+\rho_{s,y})/2$.

VBS order is detected first by measuring the bond correlations corresponding to the predicted VBS pattern (in this case, columnar).  The bond operator is simply $P_{ij}=P_{\mathbf{r}_i,\mathbf{r}_j}(\tau)$ as defined in Sec.~\ref{sec:model}.  The relevant bond correlator is thus:
\begin{equation}
\label{eqn:bondcorr}
C_V(\mathbf{r},\tau)=\frac{1}{N^2}\left[\braket{P_{\mathbf{0},\hat{x}}(0)P_{\mathbf{r},\mathbf{r}+\hat{x}}(\tau)}-\braket{P_{\mathbf{0},\hat{x}}(0)}^2\right].
\end{equation}
This does assume that the bonds will be oriented in the $x$-direction; in general, there is a degeneracy of the ground state that depends on the value of $\aniso$ (4 for $0<\aniso<1$; 6 for $\aniso=1$; and 2 for $1<\aniso<\infty$) and we measure the same correlator for $y$- and $z$-directed bonds.  We average the $x$ and $y$ correlators for $\aniso$ on the 2D side of isotropy; we average all three directions at exactly $\aniso=1$; and we only look at the $z$ correlator for $\aniso$ on the 1D side of isotropy.  Fourier transformation of this function and evaluation at $\omega=0$ and $\mathbf{k}_\mu=\pi\hat{\mu}$ (where $\mu=x,y,z$ is the relevant spatial coordinate) yields the Bragg peak corresponding to the VBS order.  This quantity diverges as the quantum volume in the thermodynamic limit. If we divide by $\beta L^2L_z$ we get the square of the order parameter for the VBS phase, which we call $\chi_V$:
\begin{equation}
\label{eqn:chiv}
\chi_V=\frac{1}{\beta L^2L_z}\int_0^\beta d\tau\sum_{\rvec}C_V(\rvec,\tau)e^{i\mathbf{k}_\mu\cdot\rvec}.
\end{equation}
Another signature of VBS order is a finite energy gap for $L\rightarrow\infty$.  This can, in principle, be extracted via a finite-size analysis of the scaling form of the specific heat at low temperatures, which should approach zero exponentially with the gap characterizing the decay rate.  In practice with our QMC method, the specific heat is notoriously difficult to measure accurately, especially for low temperatures where it is very small.  As such, we do not present any results on energy gaps in this work, but will comment on them briefly below.

If indeed a 3D photon phase did intervene between the two ordered phases, this would be detectable as a finite region of phase space wherein both order parameters, along with the energy gap of the system, vanish.  In Sec.~\ref{sec:trans}, we will detail precisely which regions of phase space we explored in search of this phase.

\section{$N-\aniso$ Phase Diagram}
\label{sec:pd}

\begin{figure}[t]
\centerline{\includegraphics[width=\columnwidth]{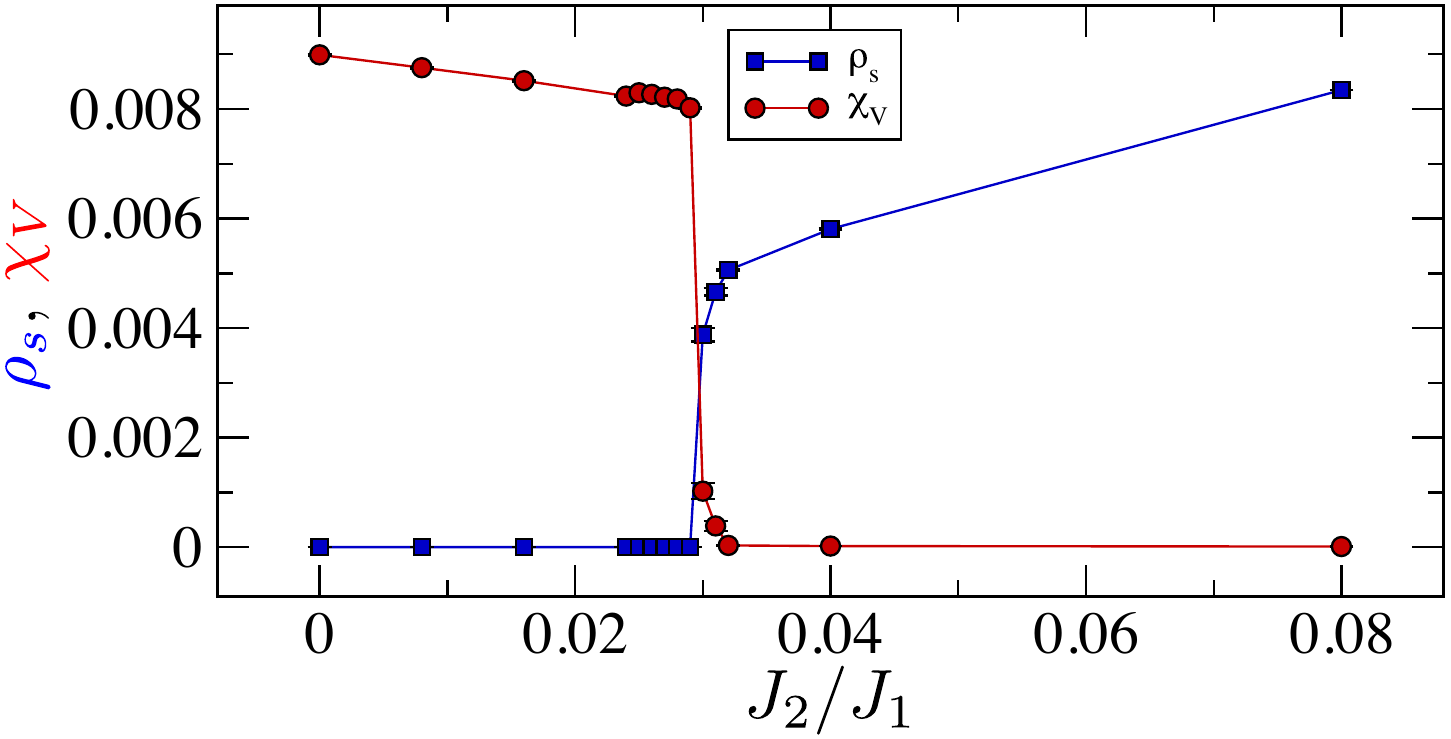}}
\caption{(color online).  Order parameters for $\aniso=1$ and $N=10$ as a function of $J_2$.  The squares mark the data for the {\Neel} order parameter; circles, data for the VBS order parameter.  In each case, the data points are obtained at the given parameters using a system with $L=L_z=16$.}
\label{fig:isoJ2}
\end{figure}

In this section, we always take $J_2=0$.  The goal here is to estimate the value of $\aniso$ for which the {\Neel}-VBS transition takes place for $3\leq N\leq9$.  We know from previous studies that in the 2D limit ($\aniso=0$), $N=4$ is {\Neel} ordered while $N=5$ is VBS ordered; in the 1D limit ($\aniso=\infty$), $N=2$ is ``{\Neel}" ordered while already at $N=3$, the ground state is VBS ordered.  Our main result here is that for the 3D isotropic case, in which we expect the minimal value of $N$ for which the VBS order onsets to be the largest, we find that $N=9$ is {\Neel} ordered while $N=10$ is VBS ordered.  Figure~\ref{fig:pd} summarizes the results of a finite size scaling analysis of both magnetic and VBS order parameters over the entire $N-\aniso$ plane.  In all cases of $\aniso$ tested, we found that the integer values of $N$ gave either {\Neel} or VBS order (see Fig.~\ref{fig:latt} for a depiction of the dimer pattern) for the ground state.  Our results at all values of $\aniso$ suggest that the transitions are first-order; but without varying $J_2$ (see the next section), we cannot be sure since $N$ can only take integer values.  The results of a detailed study of the order parameters in the isotropic case ($\aniso=1$) are shown in Fig.~\ref{fig:isoN}.

\section{\Neel-VBS Phase Transition}
\label{sec:trans}

We begin with the isotropic case, $\aniso=1$, where we have found that the {\Neel} order breaks down for $N=10$.  As shown in Fig.~\ref{fig:isoN}, there is clear VBS order for $J_2=0$.  With only a small $J_2/J_1\sim 0.028$, the {\Neel} order is restored.  In Fig.~\ref{fig:isoJ2}, we show both of the order parameters as a function of $J_2/J_1$.  Two conclusions are clear: (1) there is no intervening phase between the two ordered phases (i.e., no photon phase) and (2) the transition is first order.  Additionally, our best efforts to extract the energy gap from measurements of the specific heat for a series of low temperatures suggest that there is indeed a gap on the VBS side of the transition for $N=10$ and $J_2=0$, but due to the difficulty of accurately measuring the specific heat, as mentioned earlier, we did not pursue this study for finite $J_2$ or other values of the anisotropy parameter $\aniso$ and will not present any results on energy gaps here.

\begin{figure}[t]
\centerline{\includegraphics[width=\columnwidth]{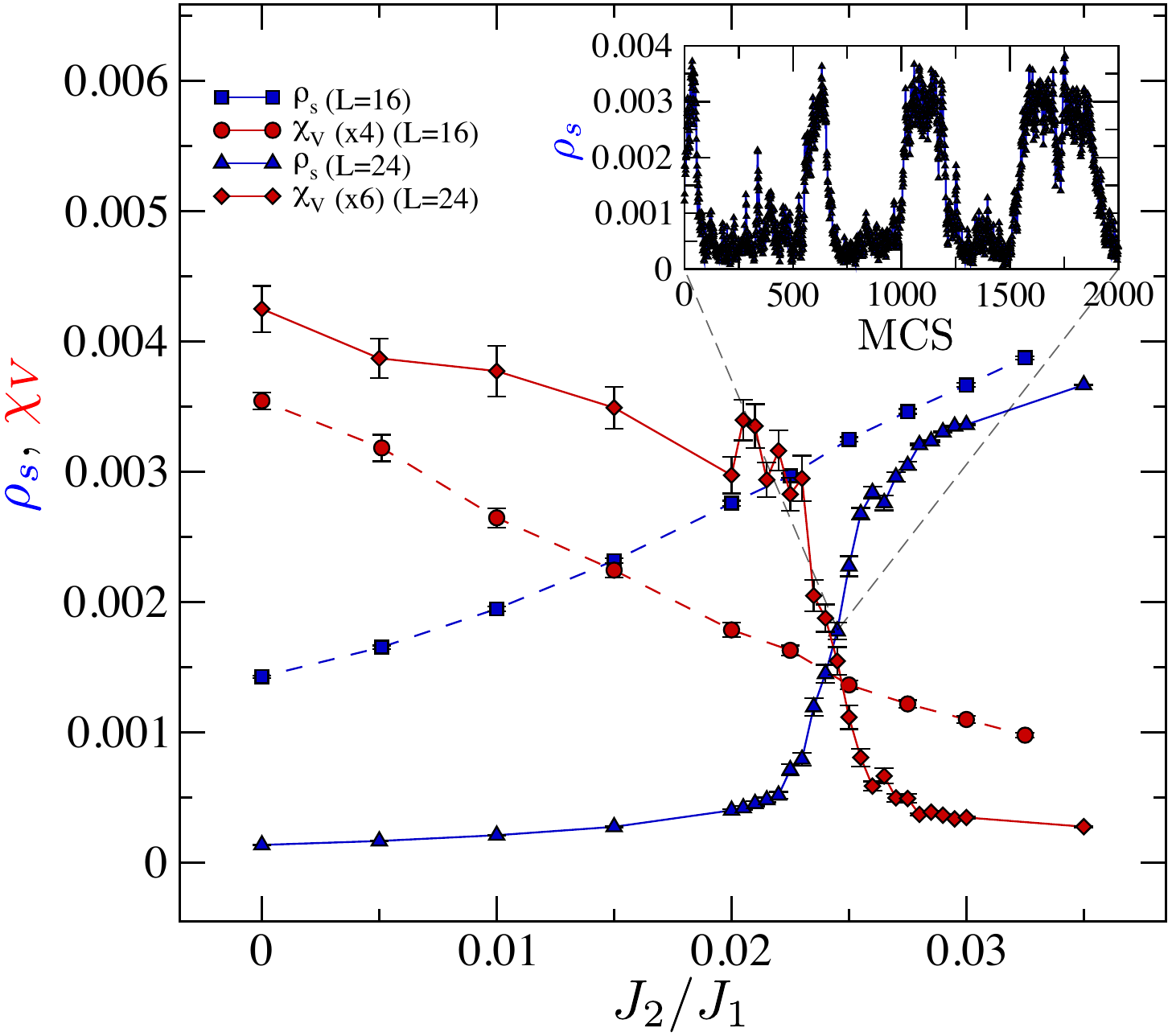}}
\caption{(color online).  Order parameters for $\aniso=0.1$ and $N=5$ as a function of $J_2/J_1$.  Two system sizes are displayed to show the trend toward a first-order transition.  The smaller system size ($16\times16\times4$) is shown with dotted lines and squares for the {\Neel} order parameter and circles for the VBS order parameter.  The solid lines correspond to the larger system size with $L=24$ and $L_z=6$.  Triangles mark the data for the {\Neel} order parameter; diamonds, data for the VBS order parameter.  The inset shows the evolution of the {\Neel} order parameter in Monte Carlo steps (each step shown actually corresponds to a bin of 500 steps) at $J_2/J_1=0.0245$.  The sharp jumps between values that correspond to each of the ordered phases provide additional evidence for the first-order nature of the transition.}
\label{fig:2DJ2}
\end{figure}

The isotropic case is, of course, very far from the scenario described in the introduction of weakly coupled planes, each in close proximity to where one expects the deconfined quantum critical point in 2D.  Therefore, we now consider $\aniso=0.1$ where $N=5$ is the minimal value for which we see VBS order, the same as in the 2D case, and this is the value we will study.  We again vary $J_2$, resulting in Fig.~\ref{fig:2DJ2}.  Again, the same two conclusions are clear: no intervening phase and a first-order transition, occurring near $J_2/J_1\sim0.025$.

There is one subtlety involved with studying the anisotropic system, which we shall now discuss.  One could argue that a sensible system geometry in three dimensions is simply a cube, and this certainly makes sense in the isotropic limit.  However, when the coupling between the planes, i.e., in the $z$ direction, is weakened (strengthened) the system becomes effectively longer (shorter) in that direction.  In the extreme cases where one is very near either the 2D or 1D limits, using a cubic system actually wastes a large amount of computer time and obtains data for a point that may be very far from the thermodynamic limit.  For example, using a $16\times16\times16$ system with $\aniso=1$ is a rather large system size perfectly capable of giving insight into the thermodynamic limit; however, at $\aniso=0.1$, this is actually a system of 16 weakly coupled planes, each with only 256 sites, far less than a typical large system size in a 2D QMC study.  The situation is even worse for, say, $\aniso=20$, where one has 256 weakly coupled chains of a paltry length of 16, which has virtually no chance of capturing large $L$ behavior.  To remedy this problem, we impose different aspect ratios of $L/L_z$ depending on the value of $\aniso$.~\cite{sandvik1999:multich}  For the value of $\aniso=0.1$, we use $L/L_z=4$.

\section{Conclusions}
\label{sec:conc}

We have presented results on the \suN antiferromagnet on the cubic lattice.  We have connected the results of the well-studied 2D and 1D cases to the 3D case and shown that there is a direct, first-order {\Neel} to VBS transition occurring between $N=9$ and $N=10$.  Additionally, we have studied a system of weakly coupled planes, each in close proximity to their deconfined critical points, and shown that while such a system is certainly a viable candidate for harboring the much sought after 3D photon phase, no evidence of the presence of such a phase exists; instead, there is again a direct, first-order transition between the two ordered phases.

Although our exploration of the phase space for the model given in Eq.~(\ref{eqn:model}) is not exhaustive, we have investigated the most likely region where one might expect to find a stable 3D photon phase.  Furthermore, we have presented compelling evidence that the continuous \Neel-VBS transition present in two dimensions immediately gives way to a first-order transition upon weak coupling of the planes.

It would be interesting to find other ways to couple layers of two-dimensional quantum critical layers that could potentially result in interesting three-dimensional spin-liquid phases. The identification of such phases in sign problem free Hamiltonians will not only allow a detailed study of the spin-liquid phase but also permit the unbiased study of the exotic quantum critical points adjacent to such phases. 
\acknowledgments

The authors thank A.~Vishwanath for stimulating discussions, A. Sandvik for collaboration on related work and K.~Beach for kindly providing numerical data used to test their code. Partial financial support was received from NSF DMR-1056536. The numerical simulations reported in the manuscript were carried out on the DLX cluster at the University of Kentucky.

\bibliography{/Users/matt/Research/UK/publications/rev_bib}

\end{document}